\documentclass[twocolumn,showpacs,preprintnumbers,amsmath,amssymb]{revtex4}

\usepackage{graphicx}
\usepackage{dcolumn}
\usepackage{bm}

\begin{document}

\title{Coherent Active-Sterile Neutrino Flavor Transformation in the Early Universe}

\author{Chad T. Kishimoto}
 \email{ckishimo@ucsd.edu}
 
\author{George M. Fuller}
 \email{gfuller@ucsd.edu}

\author{Christel J. Smith}
 \email{csmith@ucsd.edu}
\affiliation{Department of Physics, University of California, San Diego, La Jolla, CA 92093-0319}

\date{\today}

\begin{abstract}

We solve the problem of coherent Mikheyev-Smirnov-Wolfenstein (MSW) resonant active-to-sterile neutrino flavor conversion driven by an initial lepton number in the early universe.  We find incomplete destruction of lepton number in this process and a sterile neutrino energy distribution with a distinctive cusp and high energy tail. These features imply alteration of the non-zero lepton number primordial nucleosynthesis paradigm when there exist sterile neutrinos with rest masses $m_{\rm s} \sim 1\,{\rm eV}$. This could result in better light element probes of (constraints on) these particles. \end{abstract}

\maketitle

Recent advances in observational cosmology and in experimental neutrino physics promise a well constrained picture for the evolution of the early universe.  The existence of a light sterile neutrino ($\nu_s$) presents an immediate problem:  how do sterile neutrinos affect primordial seas of active neutrinos $\nu_\alpha$ or $\bar\nu_\alpha$ ($\alpha = e, \mu, \tau$) and consequentially affect the standard big bang paradigm?  In this letter we study the lepton number-driven transformation of active neutrinos to sterile neutrinos in the epoch of the early universe after weak decoupling, when neutrinos propagate coherently.  This process could leave both the active neutrinos and sterile neutrinos with distorted, non-thermal energy spectra \cite{abfw}.  A non-thermal $\nu_e$ or $\bar\nu_e$ spectrum could lead to significant modification in the relationship between lepton number and Big Bang Nucleosynthesis (BBN) $^{4}\mathrm{He}$ abundance yield \cite{abfw, man05}.  Concomitantly, a distorted $\nu_s$ distribution function changes closure mass constraints on light sterile neutrinos \cite{abfw, dod05}, allowing rest masses and vacuum mixing angles for these species in the range ($0.4\,{\rm eV} < m_{\rm s} < 5\,{\rm eV}$) suggested by the LSND experiment \cite{lsnd, lsnd1} and currently being probed by the mini-BooNE experiment \cite{miniboone}.

Active neutrinos propagating in the homogeneous early universe experience a potential stemming from forward scattering $V = 2 \sqrt{2} \zeta(3) \pi^{-2} G_F T^3 \mathcal{L}_\alpha- r_\alpha G_F^2 E_\nu T^4$, where $T$ is the photon/plasma temperature, $E_\nu$ is the neutrino energy, $r_\alpha$ is a numerical coefficient which depends on the number of relativistic charged lepton degrees of freedom and can be found in Refs.\ \cite{abfw, afp}, $G_F$ is the Fermi constant, and $\zeta\left(3\right)\approx 1.20206$. Here the potential lepton number is $\mathcal{L}_\alpha \equiv 2 L_{\nu_\alpha} + \sum_{\beta \neq \alpha} L_{\nu_\beta}$, where the individual lepton numbers are given in terms of the neutrino, antineutrino, and photon proper number densities by $L_{\nu_\alpha} \equiv (n_{\nu_\alpha} - n_{\bar\nu_\alpha}) / n_\gamma$.  Current observational bounds on these are $\vert L_{\nu_\alpha} \vert < 0.1$ \cite{dol02, ab02, wong02}, and could be slightly weaker if there are additional sources of energy density in the early universe \cite{bar03, kne01}.  We have neglected contributions to $V$ from neutrino-baryon/electron scattering since we consider relatively large lepton numbers with ${\mathcal{L}} \gg \eta$, where the baryon-to-photon ratio is $\eta \equiv {{n_{\rm b}}/{n_{\gamma}}}$ (see Refs. \cite{abfw, afp}). The second term in $V$ is negligible for the temperatures characteristic of the post weak decoupling era, $T < 3\, {\rm MeV}$.

The scattering-induced de-coherence production \cite{dol81, mck94, fv97, dib00, vol00, lee00} of seas of $\nu_s$ and $\bar\nu_s$, with rest mass $m_{\rm s} \sim 1\,{\rm eV}$, could be avoided if theses species are massless for $T > 3\,{\rm MeV}$, inflation has a low reheat temperature \cite{gel04}, or there exists a preexisting lepton number $\vert L_{\nu_\alpha} \vert > 10^{-3}$ \cite{abfw, fv95}.  However, a lepton number could subsequently, after weak decoupling, drive \cite{abfw} coherent medium-enhanced MSW \cite{ms85, w78} resonant conversion $\nu_\alpha \rightarrow \nu_s$ or $\bar\nu_\alpha \rightarrow \bar\nu_s$, depending on the sign of the lepton number.  (Resonant de-coherence production of sterile neutrinos with $m_{\rm s} \sim 1\,{\rm keV}$ with accompanying $\nu_s$ spectral distortion was considered in Refs.\ \cite{shi99,afp}.) The MSW condition for the resonant scaled neutrino energy $\epsilon = E_\nu^{\rm res} / T$ is $\delta m^2 \cos 2 \theta = 2 \epsilon T V$, or
\begin{equation}
\epsilon \mathcal{L} = \left( \frac{\delta m^2 \cos 2 \theta}{4 \sqrt{2} \zeta(3) \pi^{-2} G_F} \right) T^{-4},
\label{eq:rescondition}
\end{equation}
where $\delta m^2 \equiv m_2^2 - m_1^2$ is the difference of the squares of the vacuum neutrino mass eigenvalues.  For illustrative purposes, we consider $2 \times 2$ vacuum mixing with a one-parameter (vacuum mixing angle $\theta$) unitary transformation between weak interaction eigenstates $\vert \nu_\alpha \rangle$, $\vert \nu_s \rangle$, and energy/mass eigenstates:
\begin{eqnarray}
\vert \nu_\alpha \rangle & = & \cos \theta \vert \nu_1 \rangle + \sin \theta \vert \nu_2 \rangle ; \nonumber \\
\vert \nu_s \rangle & = & - \sin \theta \vert \nu_1 \rangle + \cos \theta \vert \nu_2 \rangle .
\end{eqnarray}

As the universe expands, the temperature falls, causing the resonance to sweep from low to higher values of the scaled neutrino energy, $\epsilon$.  This resonance sweep converts active neutrinos into sterile neutrinos, reducing $\mathcal{L}$, which accelerates the resonance sweep rate.

The evolution of $\mathcal{L}$ is dictated by the resonance sweep rate and the dimensionless adiabaticity parameter.  The adiabaticity parameter, $\gamma$, is proportional to the ratio of the width of the MSW resonance, $\delta t = \vert 1/V dV / dt \vert^{-1} \tan 2 \theta$, and the neutrino oscillation length at resonance, $L_{\text{osc}} = 4 \pi E_\nu / ( \delta m^2 \sin 2 \theta )$.  Combining the expansion rate of the universe in the radiation dominated epoch with the conservation of co-moving entropy density and the forward scattering potential $V$, the adiabaticity parameter is
\begin{eqnarray}
\gamma & \approx & \frac{\sqrt{5} \zeta^{3/4} (3)}{2^{1/8} \pi^3} \frac{(\delta m^2)^{1/4} m_{pl} G_F^{3/4}}{g^{1/2}} \frac{\sin^2 2 \theta}{\cos^{7/4} 2 \theta} \mathcal{L}^{3/4} \epsilon^{-1/4} \nonumber \\
 & & \times \left\vert 1 + \frac{\dot{g}/g}{3 H} -\frac{\dot{\mathcal{L}} / \mathcal{L}}{3 H} \right\vert^{-1} ,
\label{eq:gamma}
\end{eqnarray}
where $m_{pl}$ is the Planck mass, $g$ is the total statistical weight for relativistic species in the early universe, and $H \approx ( 4 \pi^3 / 45 )^{1/2} g^{1/2} T^2 / m_{pl}$ is the local Hubble expansion rate.  If the onset of resonant flavor conversion occurs in the epoch between weak decoupling and weak freeze out, then initially $\gamma \gg 1$ for the active-sterile mixing parameters of interest \cite{abfw}. However, when the fractional time rate of change of $\mathcal{L}$ becomes larger than the expansion rate of the universe, the evolution of the system can be non-adiabatic with $\gamma < 1$.

Large values of $\gamma$ result when many oscillation lengths fit within the resonance width.  In this case there will be a small probability of jumping from the high mass eigenstate to the low mass eigenstate.  In turn, this implies efficient flavor transformation at the MSW resonance.  Alternatively, a small value of $\gamma$ means that the resonance width is much smaller than an oscillation length, and the neutrino jumps between the two mass eigenstates, resulting in virtually no flavor transformation.  To describe intermediate cases we use the Landau-Zener jump probability, $P_{\text{LZ}} = \exp( - \pi \gamma / 2 )$ \cite{landau, zener},  which gives the likelihood for a neutrino at resonance to make the jump between mass eigenstates. It is valid in the limit where the change in $V$ across the resonance width $\delta t$ can be regarded as linear.  This is a good approximation in part because the resonance width is small compared to the causal horizon length for the values of $\theta$ and the conditions in the early universe considered here.

It follows that the evolution of the potential lepton number as the resonant scaled neutrino energy sweeps from 0 to $\epsilon$ is 
\begin{equation}
\mathcal{L} (\epsilon) = \mathcal{L}^{\text{initial}} - \frac{1}{2 \zeta(3)} \left( \frac{T_\nu}{T} \right)^3 \int_0^\epsilon \frac{x^2 (1 - e^{ - \pi \gamma(x) / 2} )}{e^{x - \eta_{\nu_\alpha}} + 1} d x,
\label{eq:Levolve}
\end{equation}
where $T_\nu$ is the temperature of the active neutrino distribution function with degeneracy parameter $\eta_{\nu_\alpha} \equiv \mu_{\nu_\alpha} / T_\nu$, and where $\mu_{\nu_\alpha}$ is the $\nu_\alpha$ chemical potential.

The evolution of the active neutrino spectrum is dictated by three conspiring factors:  the MSW resonance condition (Eq.\ \ref{eq:rescondition}), the adiabaticity parameter (Eq.\ \ref{eq:gamma}), and the evolution of potential lepton numbers through active-sterile conversion (Eq.\ \ref{eq:Levolve}).  We solve Eqs.\ (\ref{eq:rescondition}), (\ref{eq:gamma}) and (\ref{eq:Levolve}) simultaneously and self-consistently to obtain $\gamma$ and $\mathcal{L}$ as continuous functions across the entire range of $\epsilon$.

Resonant conversion of active neutrinos to sterile neutrinos begins at $\epsilon \ll 1$.  The resonance sweeps to higher values of $\epsilon$ as the temperature of the universe drops.  When $\dot{\mathcal{L}} / \mathcal{L} \ll H$, we have $\gamma \gg 1$, and adiabatic conversion of active neutrinos to sterile neutrinos ensues.  However, this trend cannot continue.  Note that the right hand side of equation (\ref{eq:rescondition}) is a monotonically increasing function of time, while the left hand side is a peaked function if one assumes continued adiabatic conversion of neutrino flavors.  At this peak, this assumption fails.  Taking the time derivative of the resonance condition, Eq.\ (\ref{eq:rescondition}), shows that the sweep rate is $\dot{\epsilon} \propto T^{-5} \dot{T} ( d (\epsilon \mathcal{L}) / d \epsilon )^{-1}$.  At the peak, $d (\epsilon \mathcal{L}) / d \epsilon = 0$, causing the sweep rate to diverge.  Taking the time derivative of both sides of Eq.\ (\ref{eq:Levolve}) and assuming that $T_\nu / T$ is constant, we conclude that $\dot{\mathcal{L}} \propto \dot{\epsilon}$.  With this relation, it follows from equation (\ref{eq:gamma}) that the MSW resonance is no longer adiabatic.  We define $\epsilon_{\text{max}}$ as the particular value of $\epsilon$ at this peak, implicitly specified by
\begin{align}
& \frac{1}{2 \zeta(3)} \frac{\epsilon_{\text{max}}^3}{e^{\epsilon_{\text{max}} - \eta_{\nu_\alpha}} + 1} \nonumber \\
 &\qquad = \mathcal{L}^{\text{initial}} - \frac{1}{2 \zeta(3)} \left( \frac{T_\nu}{T} \right)^3 \int_0^{\epsilon_{\text{max}}} \frac{x^2}{e^{x - \eta_{\nu_\alpha}} + 1} d x.
\end{align}

Our complete continuous solution for $\gamma$ shows that neutrino flavor evolution/transformation is adiabatic for $\epsilon < \epsilon_{\text{max}}$, but becomes (quickly) progressively less adiabatic for $\epsilon > \epsilon_{\text{max}}$.  For $\epsilon \geq \epsilon_{\text{max}}$, our solution yields a large, but finite resonance sweep rate, and concomitant large fractional lepton number destruction rate, $\dot{\mathcal{L}} / \mathcal{L} \gg H$, leading to $\gamma \lesssim 1$.  This behavior continues through the heart of the active neutrino distribution until the resonance sweep rate decreases to a point where $\gamma \gg 1$ again.  This last transition back to adiabatic evolution occurs at $\epsilon \sim \mathcal{O} (10)$, approximately where $\mathcal{L} (\epsilon) = 1 / (2 \zeta(3)) \epsilon^3 / (e^{\epsilon - \eta_{\nu_\alpha}} + 1)$.  (Note that this is the same condition as for $\epsilon_{\text{max}}$.)  Resonance sweep continues to higher $\epsilon$, adiabatically converting active neutrinos to sterile neutrinos.

\begin{figure}
\includegraphics[width = 2.5in, angle = 270]{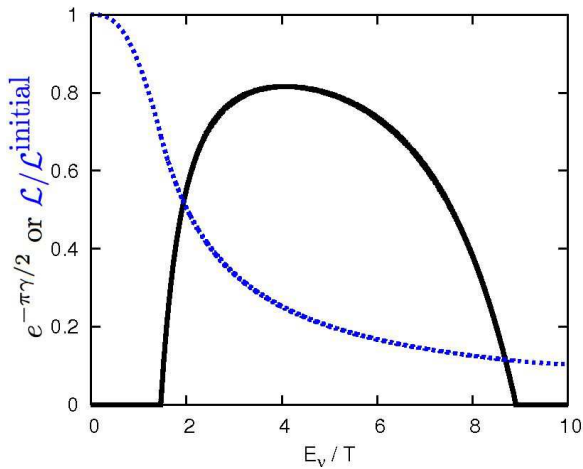}
\caption{\label{fig:gamma}Landau-Zener jump probability $e^{-\pi \gamma / 2}$ (solid curve) and potential lepton number given as a fraction of its initial value (dashed curve) are shown as a function of MSW scaled resonance energy $E_\nu / T$.  Here we assume $\delta m^2 = 1\,{\rm eV}^2$, $\sin^2 2 \theta = 10^{-3}$, and initial individual lepton numbers $L_{\nu_\mu} = L_{\nu_\tau} = 0.15$ and $L_{\nu_e} = 0.0343$.}
\end{figure}

The evolution of the Landau-Zener jump probability $e^{- \pi \gamma / 2}$ and the history of the potential lepton number as  a fraction of its initial value are both shown in Figure \ref{fig:gamma} for the particular case where $\delta m^2 = 1\,{\rm eV}^2$, $\sin^2 2 \theta = 10^{-3}$, and where we assume initial lepton numbers near their conventional upper limits, $L_{\nu_\mu} = L_{\nu_\tau} = 0.15$ and $L_{\nu_e} = 0.0343$.  For this particular case $\epsilon_{\text{max}} = 1.46$, and Figure \ref{fig:gamma} shows the rather abrupt (but continuous) change to non-adiabatic evolution for $\epsilon \approx \epsilon_{\text{max}}$.  In this example, the final transition back to adiabatic evolution occurs at $\epsilon \approx 8.9$.  Altogether, more than $90 \%$ of the initial potential lepton number is destroyed for this case.  We find that the fractional depletion of potential lepton number is $\sim 90 \%$ across a wide range of initial values of this parameter.  This, in turn, suggests that this new solution will result in little change in existing closure mass constraints on light sterile neutrinos \cite{abfw}.

\begin{figure}
\includegraphics[width = 2.5in, angle = 270]{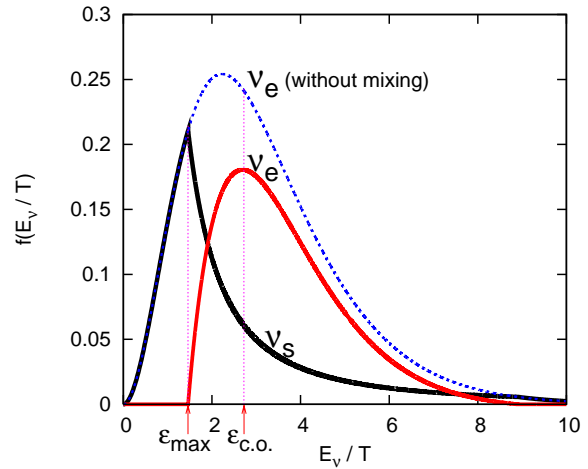}
\caption{\label{fig:dist}Shown are the original $\nu_e$ distribution function (dashed curve), the final $\nu_e$ distribution function (lighter solid curve), and final $\nu_s$ distribution function (heavier solid curve) all as functions of scaled neutrino energy $E_\nu / T$ for a $\nu_e \rightarrow \nu_s$ resonant, coherent flavor conversion process with $\delta m^2 = 1\,{\rm eV}^2$, $\sin^2 2 \theta = 10^{-3}$, and individual lepton numbers as in Fig. \ref{fig:gamma}.  Vertical dotted lines indicate $\epsilon_{\text{max}}$ and $\epsilon_{\text{c.o.}}$.}
\end{figure}

Figure \ref{fig:dist} shows the original $\nu_e$ Fermi-Dirac ($f (E_\nu / T) = 1/ [ T_\nu^3 F_2 (\eta_\nu) ] E_\nu^2 / (e^{E_\nu / T_\nu - \eta_\nu} + 1)$, where $F_2 (\eta_\nu) \equiv \int_0^\infty x^2 / (e^{x - \eta_\nu} + 1) dx$) and final $\nu_e$ and the $\nu_s$ energy distribution functions resulting from $\nu_e \rightarrow \nu_s$ resonance sweep for the example parameters of Fig. \ref{fig:gamma}.  Forced, adiabatic resonance sweep to $\epsilon_{\text{c.o.}}$ would result in complete depletion of the initial potential lepton number.  $\epsilon_{\text{max}}$ and $\epsilon_{\text{c.o.}}$ are shown for this case in Fig. \ref{fig:dist}.  Forced, adiabatic resonance sweep would result in a final $\nu_e$ spectrum identical to the initial one except cut-off (hence, \lq\lq $\text{c.o.}$\rq\rq), with zero population, for $E_\nu / T \leq \epsilon_{\text{c.o.}}$.  The $\nu_s$ distribution in this case would be simply the complement.  By contrast, with the full resonance sweep solution presented here we see that the actual final $\nu_e$ spectrum has a population deficit relative to the original distribution,  even for $E_\nu / T > \epsilon_{\text{c.o.}}$.  Likewise, the actual final $\nu_s$ spectrum will now have a tail extending to higher $E_\nu/T$.  Including simultaneous active-sterile and active-active neutrino flavor transformation in a full $4 \times 4$ scheme will modify this result, but we can expect some general features of our solution to remain.  In particular, although neutrino flavor evolution will start out adiabatic, the transition to non-adiabatic evolution could be altered by, {\it e.g.}, active-active neutrino mixing partially ``filling-in'' depleted $\nu_e$ population \cite{abfw}.

The BBN $^4\text{He}$ yield can depend sensitively on the shape of the $\nu_e$ energy distribution function \cite{kir03, asf, abfw}.  This is because the neutron-to-proton ratio $n / p$ is a crucial determinant of the $^4\text{He}$ abundance and, in turn, this ratio is set by the competition among the charged current weak neutron/proton interconversion processes:
\begin{eqnarray}
\nu_e + n & \rightleftharpoons & p + e^- ; \nonumber \\
\bar\nu_e + p & \rightleftharpoons & n + e^+ ; \\
n & \rightleftharpoons & p + e^- + \bar\nu_e. \nonumber
\end{eqnarray}
The net rate for the forward direction in the first of these processes will be reduced if $\nu_e$-population is removed via $\nu_e \rightarrow \nu_s$, resulting in a larger $n / p$ and, hence, a larger $^4\text{He}$ yield.  Likewise, a negative potential lepton number-driven $\bar\nu_e \rightarrow \bar\nu_s$ scenario could result in $\bar\nu_e$ spectral depletion which will result in a smaller $n / p$ and, hence, less $^4\text{He}$.  Removing $\nu_e$ ($\bar\nu_e$) population at higher $E_\nu / T$ values in the energy distribution function accentuates these effects because the cross section for the $\nu_e$  ($\bar\nu_e$) capture process scales as $E_\nu^2$ and because the Fermi-Dirac spectral peak, where neutrino populations are large, corresponds to values of neutrino energy satisfying $E_\nu / T > \epsilon_{\text{c.o.}}$ for the potential lepton numbers $\mathcal{L}$ of interest here.  As a consequence, our full resonance sweep scenario can result in significant alteration in $^4\text{He}$ yield over the forced, adiabatic scenario.

\begin{figure}
\includegraphics[width = 2.5in, angle = 270]{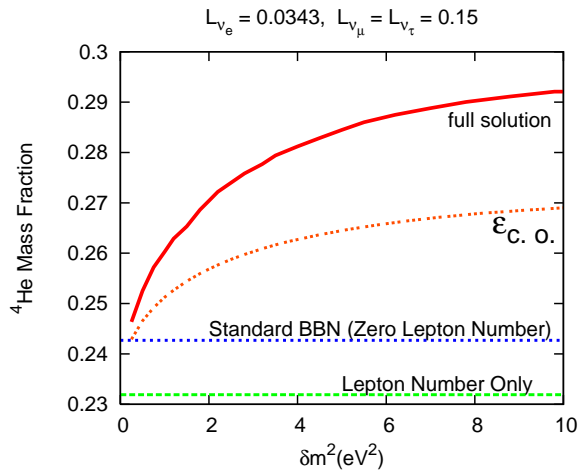}
\caption{\label{fig:he4}Primordial nucleosynthesis (BBN) $^4\text{He}$ abundance yield as a function of $\delta m^2$ for the $\nu_e \rightarrow \nu_s$ channel and the indicated initial individual lepton numbers (same as in Fig. \ref{fig:gamma}).  Standard BBN (zero lepton numbers, no sterile neutrinos) is the heavy dashed horizontal line.  The case for BBN with the indicated lepton number, but no active-sterile mixing is the light dashed horizontal line.  The case for forced, adiabatic resonance sweep to $\epsilon_{\text{c.o.}}$ is the light dotted line.  The full non-adiabatic solution is given by the heavy solid line.}
\end{figure}

We have computed the BBN $^4\text{He}$ abundance yield with a version of the Kawano-Wagoner-Fowler-Hoyle code \cite{kaw93,whf} modified to allow for dynamic alteration/distortion in the neutrino energy distribution functions.  The results of these calculations for the initial lepton numbers adopted in the example of Fig.\ \ref{fig:gamma} are shown in Figure \ref{fig:he4}.  The standard (zero lepton number, no sterile neutrinos) BBN $^4\text{He}$ abundance yield mass fraction is $\approx 24\%$ when we adopt neutron lifetime $\tau_n = 887.8\,{\rm s}$ and $\eta = 6.1102 \times 10^{-10}$.  The adopted value of $\eta$ corresponds to the central value of the cosmic microwave background radiation acoustic peak-determined WMAP 3-year data, $\eta = (6.11 \pm 0.22) \times 10^{-10}$ \cite{wmap3}. The observational error in $\eta$ corresponds to a $\pm 0.03 \%$ range in the calculated $^4\text{He}$ abundance yield. 

Alternatively, the case with the example lepton numbers but with no active-sterile neutrino mixing gives a healthy $^4\text{He}$ yield suppression.  However, once the spectral distortion is included the $^4\text{He}$ yield is larger than in standard BBN.  Given that the observationally-inferred helium abundance is between $23 \% - 26 \%$ \cite{olive04} (and possibly more precisely determined \cite{olive97, izo04}), we see that the dramatically larger $^4\text{He}$ yield in the cases with $\nu_e$ spectral distortion may allow for new constraints on a combination of lepton number and sterile neutrino masses.  Our resonance sweep solution gives a larger $^4\text{He}$ yield than in previous models of active-sterile neutrino transformation which employ, {\it e.g.}, forced, adiabatic resonance sweep to $\epsilon_{\rm c.o.}$ \cite{abfw}.  This shows the sensitivity of BBN abundance yields to sterile neutrino-induced active neutrino spectral distortion.  This effect eventually may allow light element probes/constraints on the sterile neutrino sector which complement those of mini-BooNE and may extend to sterile neutrino mass/mixing parameters currently inaccessible experimentally.

We would like to thank K. Abazajian, N. Bell, M. Smith, and especially M. Patel for insightful discussions.  This work was supported in part by NSF Grant PHY-04-00359 at UCSD.  C.T.K. would like to acknowledge a fellowship from the ARCS Foundation, Inc.

\bibliography{ressweep}

\begin{thebibliography}{33}
\expandafter\ifx\csname natexlab\endcsname\relax\def\natexlab#1{#1}\fi
\expandafter\ifx\csname bibnamefont\endcsname\relax
  \def\bibnamefont#1{#1}\fi
\expandafter\ifx\csname bibfnamefont\endcsname\relax
  \def\bibfnamefont#1{#1}\fi
\expandafter\ifx\csname citenamefont\endcsname\relax
  \def\citenamefont#1{#1}\fi
\expandafter\ifx\csname url\endcsname\relax
  \def\url#1{\texttt{#1}}\fi
\expandafter\ifx\csname urlprefix\endcsname\relax\def\urlprefix{URL }\fi
\providecommand{\bibinfo}[2]{#2}
\providecommand{\eprint}[2][]{\url{#2}}

\bibitem[{\citenamefont{Abazajian et~al.}(2005)\citenamefont{Abazajian, Bell,
  Fuller, and Wong}}]{abfw}
\bibinfo{author}{\bibfnamefont{K.}~\bibnamefont{Abazajian}},
  \bibinfo{author}{\bibfnamefont{N.~F.} \bibnamefont{Bell}},
  \bibinfo{author}{\bibfnamefont{G.~M.} \bibnamefont{Fuller}},
  \bibnamefont{and} \bibinfo{author}{\bibfnamefont{Y.~Y.~Y.}
  \bibnamefont{Wong}}, \bibinfo{journal}{Phys.\ Rev.\ D}
  \textbf{\bibinfo{volume}{72}}, \bibinfo{pages}{063004}
  (\bibinfo{year}{2005}).

\bibitem[{\citenamefont{Mangano et~al.}(2005)\citenamefont{Mangano, Miele,
  Pastor, Pinto, Pisanti, and Serpico}}]{man05}
\bibinfo{author}{\bibfnamefont{G.}~\bibnamefont{Mangano}},
  \bibinfo{author}{\bibfnamefont{G.}~\bibnamefont{Miele}},
  \bibinfo{author}{\bibfnamefont{S.}~\bibnamefont{Pastor}},
  \bibinfo{author}{\bibfnamefont{T.}~\bibnamefont{Pinto}},
  \bibinfo{author}{\bibfnamefont{O.}~\bibnamefont{Pisanti}}, \bibnamefont{and}
  \bibinfo{author}{\bibfnamefont{P.~D.} \bibnamefont{Serpico}},
  \bibinfo{journal}{Nucl.\ Phys.\ B} \textbf{\bibinfo{volume}{729}},
  \bibinfo{pages}{221} (\bibinfo{year}{2005}).

\bibitem[{\citenamefont{Dodelson et~al.}(2005)\citenamefont{Dodelson,
  Melchiorri, and Slosar}}]{dod05}
\bibinfo{author}{\bibfnamefont{S.}~\bibnamefont{Dodelson}},
  \bibinfo{author}{\bibfnamefont{A.}~\bibnamefont{Melchiorri}},
  \bibnamefont{and} \bibinfo{author}{\bibfnamefont{A.}~\bibnamefont{Slosar}}
  (\bibinfo{year}{2005}), \eprint{astro-ph/0511500}.

\bibitem[{\citenamefont{Eitel}(2000)}]{lsnd}
\bibinfo{author}{\bibfnamefont{K.}~\bibnamefont{Eitel}}, \bibinfo{journal}{New
  J.\ Phys.} \textbf{\bibinfo{volume}{2}}, \bibinfo{pages}{1}
  (\bibinfo{year}{2000}).

\bibitem[{\citenamefont{Athanassopoulos et~al.}(1995)}]{lsnd1}
\bibinfo{author}{\bibfnamefont{C.}~\bibnamefont{Athanassopoulos}}
  \bibnamefont{et~al.}, \bibinfo{journal}{Phys.\ Rev.\ Lett.}
  \textbf{\bibinfo{volume}{75}}, \bibinfo{pages}{2650} (\bibinfo{year}{1995}).

\bibitem[{\citenamefont{McGregor}(2003)}]{miniboone}
\bibinfo{author}{\bibfnamefont{G.}~\bibnamefont{McGregor}}, in
  \emph{\bibinfo{booktitle}{Particle Physics and Cosmology: Third Tropical
  Workshop on Particle Physics and Cosmology - Neutrinos, Branes, and
  Cosmology}}, edited by \bibinfo{editor}{\bibfnamefont{J.~F.}
  \bibnamefont{Nieves}} \bibnamefont{and} \bibinfo{editor}{\bibfnamefont{C.~N.}
  \bibnamefont{Leung}} (\bibinfo{publisher}{AIP}, \bibinfo{address}{New York},
  \bibinfo{year}{2003}), AIP Conf.\ Proc.\ No.\ 655, p.~\bibinfo{pages}{58}.

\bibitem[{\citenamefont{Abazajian et~al.}(2001)\citenamefont{Abazajian, Fuller,
  and Patel}}]{afp}
\bibinfo{author}{\bibfnamefont{K.}~\bibnamefont{Abazajian}},
  \bibinfo{author}{\bibfnamefont{G.~M.} \bibnamefont{Fuller}},
  \bibnamefont{and} \bibinfo{author}{\bibfnamefont{M.}~\bibnamefont{Patel}},
  \bibinfo{journal}{Phys.\ Rev.\ D} \textbf{\bibinfo{volume}{64}},
  \bibinfo{pages}{023501} (\bibinfo{year}{2001}).

\bibitem[{\citenamefont{Dolgov et~al.}(2002)\citenamefont{Dolgov, Hansen,
  Pastor, Petcov, Raffelt, and Semikoz}}]{dol02}
\bibinfo{author}{\bibfnamefont{A.~D.} \bibnamefont{Dolgov}},
  \bibinfo{author}{\bibfnamefont{S.~H.} \bibnamefont{Hansen}},
  \bibinfo{author}{\bibfnamefont{S.}~\bibnamefont{Pastor}},
  \bibinfo{author}{\bibfnamefont{S.~T.} \bibnamefont{Petcov}},
  \bibinfo{author}{\bibfnamefont{G.~G.} \bibnamefont{Raffelt}},
  \bibnamefont{and} \bibinfo{author}{\bibfnamefont{D.~V.}
  \bibnamefont{Semikoz}}, \bibinfo{journal}{Nucl.\ Phys.\ B}
  \textbf{\bibinfo{volume}{632}}, \bibinfo{pages}{363} (\bibinfo{year}{2002}).

\bibitem[{\citenamefont{Abazajian et~al.}(2002)\citenamefont{Abazajian, Beacom,
  and Bell}}]{ab02}
\bibinfo{author}{\bibfnamefont{K.~N.} \bibnamefont{Abazajian}},
  \bibinfo{author}{\bibfnamefont{J.~F.} \bibnamefont{Beacom}},
  \bibnamefont{and} \bibinfo{author}{\bibfnamefont{N.~F.} \bibnamefont{Bell}},
  \bibinfo{journal}{Phys.\ Rev.\ D} \textbf{\bibinfo{volume}{66}},
  \bibinfo{pages}{013008} (\bibinfo{year}{2002}).

\bibitem[{\citenamefont{Wong}(2002)}]{wong02}
\bibinfo{author}{\bibfnamefont{Y.~Y.~Y.} \bibnamefont{Wong}},
  \bibinfo{journal}{Phys.\ Rev.\ D} \textbf{\bibinfo{volume}{66}},
  \bibinfo{pages}{025015} (\bibinfo{year}{2002}).

\bibitem[{\citenamefont{Barger et~al.}(2003)\citenamefont{Barger, Kneller,
  Langacker, Marfatia, and Steigman}}]{bar03}
\bibinfo{author}{\bibfnamefont{V.}~\bibnamefont{Barger}},
  \bibinfo{author}{\bibfnamefont{J.~P.} \bibnamefont{Kneller}},
  \bibinfo{author}{\bibfnamefont{P.}~\bibnamefont{Langacker}},
  \bibinfo{author}{\bibfnamefont{D.}~\bibnamefont{Marfatia}}, \bibnamefont{and}
  \bibinfo{author}{\bibfnamefont{G.}~\bibnamefont{Steigman}},
  \bibinfo{journal}{Phys.\ Lett.\ B} \textbf{\bibinfo{volume}{569}},
  \bibinfo{pages}{123} (\bibinfo{year}{2003}).

\bibitem[{\citenamefont{Kneller et~al.}(2001)\citenamefont{Kneller, Scherrer,
  Steigman, and Walker}}]{kne01}
\bibinfo{author}{\bibfnamefont{J.~P.} \bibnamefont{Kneller}},
  \bibinfo{author}{\bibfnamefont{R.~J.} \bibnamefont{Scherrer}},
  \bibinfo{author}{\bibfnamefont{G.}~\bibnamefont{Steigman}}, \bibnamefont{and}
  \bibinfo{author}{\bibfnamefont{T.~P.} \bibnamefont{Walker}},
  \bibinfo{journal}{Phys.\ Rev.\ D} \textbf{\bibinfo{volume}{64}},
  \bibinfo{pages}{123506} (\bibinfo{year}{2001}).

\bibitem[{\citenamefont{Volkas and Wong}(2000)}]{vol00}
\bibinfo{author}{\bibfnamefont{R.~R.} \bibnamefont{Volkas}} \bibnamefont{and}
  \bibinfo{author}{\bibfnamefont{Y.~Y.~Y.} \bibnamefont{Wong}},
  \bibinfo{journal}{Phys.\ Rev.\ D} \textbf{\bibinfo{volume}{62}},
  \bibinfo{pages}{093024} (\bibinfo{year}{2000}).

\bibitem[{\citenamefont{Lee et~al.}(2000)\citenamefont{Lee, Volkas, and
  Wong}}]{lee00}
\bibinfo{author}{\bibfnamefont{K.~S.~M.} \bibnamefont{Lee}},
  \bibinfo{author}{\bibfnamefont{R.~R.} \bibnamefont{Volkas}},
  \bibnamefont{and} \bibinfo{author}{\bibfnamefont{Y.~Y.~Y.}
  \bibnamefont{Wong}}, \bibinfo{journal}{Phys.\ Rev.\ D}
  \textbf{\bibinfo{volume}{62}}, \bibinfo{pages}{093025}
  (\bibinfo{year}{2000}).

\bibitem[{\citenamefont{Dolgov}(1981)}]{dol81}
\bibinfo{author}{\bibfnamefont{A.~D.} \bibnamefont{Dolgov}},
  \bibinfo{journal}{Yad.\ Fiz.} \textbf{\bibinfo{volume}{33}},
  \bibinfo{pages}{1309} (\bibinfo{year}{1981}).

\bibitem[{\citenamefont{McKellar and Thomson}(1994)}]{mck94}
\bibinfo{author}{\bibfnamefont{B.~H.~J.} \bibnamefont{McKellar}}
  \bibnamefont{and} \bibinfo{author}{\bibfnamefont{M.~J.}
  \bibnamefont{Thomson}}, \bibinfo{journal}{Phys.\ Rev.\ D}
  \textbf{\bibinfo{volume}{49}}, \bibinfo{pages}{2710} (\bibinfo{year}{1994}).

\bibitem[{\citenamefont{Foot and Volkas}(1997)}]{fv97}
\bibinfo{author}{\bibfnamefont{R.}~\bibnamefont{Foot}} \bibnamefont{and}
  \bibinfo{author}{\bibfnamefont{R.~R.} \bibnamefont{Volkas}},
  \bibinfo{journal}{Phys.\ Rev.\ D} \textbf{\bibinfo{volume}{55}},
  \bibinfo{pages}{5147} (\bibinfo{year}{1997}).

\bibitem[{\citenamefont{{Di Bari} et~al.}(2000)\citenamefont{{Di Bari}, Lipari,
  and Lusignoli}}]{dib00}
\bibinfo{author}{\bibfnamefont{P.}~\bibnamefont{{Di Bari}}},
  \bibinfo{author}{\bibfnamefont{P.}~\bibnamefont{Lipari}}, \bibnamefont{and}
  \bibinfo{author}{\bibfnamefont{M.}~\bibnamefont{Lusignoli}},
  \bibinfo{journal}{Int.\ J.\ Mod.\ Phys.} \textbf{\bibinfo{volume}{A15}},
  \bibinfo{pages}{2289} (\bibinfo{year}{2000}).

\bibitem[{\citenamefont{Gelmini et~al.}(2004)\citenamefont{Gelmini,
  Palomares-Ruiz, and Pascoli}}]{gel04}
\bibinfo{author}{\bibfnamefont{G.}~\bibnamefont{Gelmini}},
  \bibinfo{author}{\bibfnamefont{S.}~\bibnamefont{Palomares-Ruiz}},
  \bibnamefont{and} \bibinfo{author}{\bibfnamefont{S.}~\bibnamefont{Pascoli}},
  \bibinfo{journal}{Phys.\ Rev.\ Lett.} \textbf{\bibinfo{volume}{93}},
  \bibinfo{pages}{081302} (\bibinfo{year}{2004}).

\bibitem[{\citenamefont{Foot and Volkas}(1995)}]{fv95}
\bibinfo{author}{\bibfnamefont{R.}~\bibnamefont{Foot}} \bibnamefont{and}
  \bibinfo{author}{\bibfnamefont{R.~R.} \bibnamefont{Volkas}},
  \bibinfo{journal}{Phys.\ Rev.\ Lett.} \textbf{\bibinfo{volume}{75}},
  \bibinfo{pages}{4350} (\bibinfo{year}{1995}).

\bibitem[{\citenamefont{Mikheyev and Smirnov}(1985)}]{ms85}
\bibinfo{author}{\bibfnamefont{S.~P.} \bibnamefont{Mikheyev}} \bibnamefont{and}
  \bibinfo{author}{\bibfnamefont{A.~Y.} \bibnamefont{Smirnov}},
  \bibinfo{journal}{Yad.\ Fiz.} \textbf{\bibinfo{volume}{42}},
  \bibinfo{pages}{1441} (\bibinfo{year}{1985}).

\bibitem[{\citenamefont{Wolfenstein}(1978)}]{w78}
\bibinfo{author}{\bibfnamefont{L.}~\bibnamefont{Wolfenstein}},
  \bibinfo{journal}{Phys.\ Rev.\ D} \textbf{\bibinfo{volume}{17}},
  \bibinfo{pages}{2369} (\bibinfo{year}{1978}).

\bibitem[{\citenamefont{Shi and Fuller}(1999)}]{shi99}
\bibinfo{author}{\bibfnamefont{X.}~\bibnamefont{Shi}} \bibnamefont{and}
  \bibinfo{author}{\bibfnamefont{G.~M.} \bibnamefont{Fuller}},
  \bibinfo{journal}{Phys.\ Rev.\ Lett.} \textbf{\bibinfo{volume}{82}},
  \bibinfo{pages}{2832} (\bibinfo{year}{1999}).

\bibitem[{\citenamefont{Landau}(1932)}]{landau}
\bibinfo{author}{\bibfnamefont{L.~D.} \bibnamefont{Landau}},
  \bibinfo{journal}{Phys.\ Z.\ Sowjetunion} \textbf{\bibinfo{volume}{2}},
  \bibinfo{pages}{46} (\bibinfo{year}{1932}).

\bibitem[{\citenamefont{Zener}(1932)}]{zener}
\bibinfo{author}{\bibfnamefont{C.}~\bibnamefont{Zener}},
  \bibinfo{journal}{Proc.\ R.\ Soc.\ London Ser.\ A}
  \textbf{\bibinfo{volume}{137}}, \bibinfo{pages}{696} (\bibinfo{year}{1932}).

\bibitem[{\citenamefont{Kirilova}(2003)}]{kir03}
\bibinfo{author}{\bibfnamefont{D.~P.} \bibnamefont{Kirilova}},
  \bibinfo{journal}{Astropart.\ Phys.} \textbf{\bibinfo{volume}{19}},
  \bibinfo{pages}{409} (\bibinfo{year}{2003}).

\bibitem[{\citenamefont{Abazajian et~al.}(1999)\citenamefont{Abazajian, Shi,
  and Fuller}}]{asf}
\bibinfo{author}{\bibfnamefont{K.}~\bibnamefont{Abazajian}},
  \bibinfo{author}{\bibfnamefont{X.}~\bibnamefont{Shi}}, \bibnamefont{and}
  \bibinfo{author}{\bibfnamefont{G.~M.} \bibnamefont{Fuller}}
  (\bibinfo{year}{1999}), \eprint{astro-ph/9909320}.

\bibitem[{\citenamefont{Smith et~al.}(1993)\citenamefont{Smith, Kawano, and
  Malaney}}]{kaw93}
\bibinfo{author}{\bibfnamefont{M.~S.} \bibnamefont{Smith}},
  \bibinfo{author}{\bibfnamefont{L.~H.} \bibnamefont{Kawano}},
  \bibnamefont{and} \bibinfo{author}{\bibfnamefont{R.~A.}
  \bibnamefont{Malaney}}, \bibinfo{journal}{Astrophys.\ J.\ Suppl.}
  \textbf{\bibinfo{volume}{85}}, \bibinfo{pages}{219} (\bibinfo{year}{1993}).

\bibitem[{\citenamefont{Wagoner et~al.}(1967)\citenamefont{Wagoner, Fowler, and
  Hoyle}}]{whf}
\bibinfo{author}{\bibfnamefont{R.~V.} \bibnamefont{Wagoner}},
  \bibinfo{author}{\bibfnamefont{W.~A.} \bibnamefont{Fowler}},
  \bibnamefont{and} \bibinfo{author}{\bibfnamefont{F.}~\bibnamefont{Hoyle}},
  \bibinfo{journal}{Astrophys.\ J.} \textbf{\bibinfo{volume}{148}},
  \bibinfo{pages}{3} (\bibinfo{year}{1967}).

\bibitem[{\citenamefont{Spergel et~al.}(2006)}]{wmap3}
\bibinfo{author}{\bibfnamefont{D.~N.} \bibnamefont{Spergel}}
  \bibnamefont{et~al.} (\bibinfo{year}{2006}), \eprint{astro-ph/0603449}.

\bibitem[{\citenamefont{Olive and Skillman}(2004)}]{olive04}
\bibinfo{author}{\bibfnamefont{K.~A.} \bibnamefont{Olive}} \bibnamefont{and}
  \bibinfo{author}{\bibfnamefont{E.~D.} \bibnamefont{Skillman}},
  \bibinfo{journal}{Astrophys.\ J.} \textbf{\bibinfo{volume}{617}},
  \bibinfo{pages}{29} (\bibinfo{year}{2004}).

\bibitem[{\citenamefont{Olive et~al.}(1997)\citenamefont{Olive, Skillman, and
  Steigman}}]{olive97}
\bibinfo{author}{\bibfnamefont{K.~A.} \bibnamefont{Olive}},
  \bibinfo{author}{\bibfnamefont{E.}~\bibnamefont{Skillman}}, \bibnamefont{and}
  \bibinfo{author}{\bibfnamefont{G.}~\bibnamefont{Steigman}},
  \bibinfo{journal}{Astrophys.\ J.} \textbf{\bibinfo{volume}{483}},
  \bibinfo{pages}{788} (\bibinfo{year}{1997}).

\bibitem[{\citenamefont{Izotov and Thuan}(2004)}]{izo04}
\bibinfo{author}{\bibfnamefont{Y.~I.} \bibnamefont{Izotov}} \bibnamefont{and}
  \bibinfo{author}{\bibfnamefont{T.~X.} \bibnamefont{Thuan}},
  \bibinfo{journal}{Astrophys.\ J.} \textbf{\bibinfo{volume}{602}},
  \bibinfo{pages}{200} (\bibinfo{year}{2004}).

\end{thebibliography}

\end{document}